\documentclass[prl ,showpacs,showkeys,twocolumn]{revtex4-1}

\usepackage{graphicx}
\usepackage{latexsym}
\usepackage{amsmath}
\usepackage{amssymb}
\usepackage{amsfonts}
\usepackage{color}
\usepackage[dvipsnames]{xcolor}
\usepackage{textcomp}
\usepackage[normalem]{ulem}

\newcommand{\xx}{\mathbf{x}}
\newcommand{\kk}{\mathbf{k}}
\newcommand{\Meff}{M_\textrm{eff}}

\begin{document}


\title{Vortex unbinding transition in nonequilibrium photon condensates}

\author{Vladimir N. Gladilin}
\author{Michiel Wouters}

\affiliation{TQC, Universiteit Antwerpen, Universiteitsplein 1,
B-2610 Antwerpen, Belgium}

\date{\today}

\begin{abstract}
 We present a theoretical study of a
Berezinskii-Kosterlitz-Thouless like phase transition in lattices of
nonequilibrium photon condensates. Starting from linearized
fluctuation theory and the properties of vortices, we propose an
analytical formula for the critical point containing four fitting
parameters, that captures well all our numerical simulations. We
find that the ordered phase becomes more stable when driving and
dissipation is increased.
\end{abstract}

\maketitle

{\em Introduction --} {Thermalization of cavity photons by repeated
absorption and emission by dye molecules \cite{klaers_therm} has led
to photonic Bose-Einstein condensates
\cite{klaers10,marelic16,greveling18} and presents an invitation to
study photonic systems from a quantum fluid perspective
\cite{carusotto13}. }  The dimensional reduction implied by the
microcavity structure, required to give the photon a nonzero rest
mass, immediately raises the issue of the Hohenberg-Mermin-Wagner
theorem that forbids BEC in two dimensions at finite temperature
\cite{bec_book}. Experimentally, this has been avoided by a harmonic
trapping potential, that modifies the density of states, and
condensation in the ground state has been
observed~\cite{klaers10,marelic16,greveling18}. In this geometry,
the condensate is however not spatially extended. 

In extended two-dimensional interacting bose gases, there is a
Berezinskii-Kosterlitz-Thouless (BKT) transition
\cite{bere71,KT73,bec_book,prokofev01} that separates a phase with
free vortices from a phase where all vortex-antivortex pairs are
bound. For this phase transition to occur, interactions are crucial:
in their absence, the vortex core size tends to infinity and
vortices cease to be well defined excitations. At first sight, this
seems to be fatal for the possibility of a phase transition in
extended photon condensates that show negligible interactions except
for a very slow thermal nonlinearity \cite{alaeian17}.

For photon condensates, the main deviation from the ideal bose gas
comes from driving and dissipation. The confining mirrors are never
perfect so that photon losses have to be compensated by continuous
pumping of the dye molecules. We have shown recently that the
nonlinear dynamics of driving and dissipation renders the vortex
core size in a lattice of coupled photon condenstates finite
\cite{gladilin20b}. Motivated by experiments on exciton-polaritons,
it has already been numerically demonstrated that nonequilibrium
interacting bose gases feature a BKT-like transition
\cite{dagva15,caputo2016,gladilin19}. At the same time it has been
shown that the phase dynamics is actually in the Kardar-Parisi-Zhang
(KPZ) universality class, which has been argued to destroy the
superfluid-like phase \cite{altman15,wachtel,sieberer}. In practice,
however, the KPZ physics can be limited to very large system sizes,
so that in experimental systems the BKT-like physics dominates
\cite{dagva15,caputo2016,gladilin19,comaron18}.

We will show in this Letter that numerical classical field
simulations on a finite lattice of photon condensates predict a
BKT-like transition between states with and without free
vortex-antivortex pairs. In addition, we will construct an
analytical expression containing a few fitting constants that
explains the numerically obtained dependence of the critical point
on the system parameters.

{\em Model --} The simplest theoretical model to describe a lattice
of coupled photonic cavities in the quantum degenerate regime is the
generalized Gross-Pitaevskii equation (gGPE) \cite{gladilin20}:
\begin{align}
i \hbar \frac{\partial \psi(\xx,t)}{\partial t}  = \frac{i}{2}\left[
B_{21}M_2(\xx,t) - B_{12}
M_1(\xx,t)-\gamma\right]\psi(\xx,t) \nonumber \\
- (1-i\kappa) J  \sum_{\xx' \in \mathcal N_\xx} \psi(\xx',t) +
\sqrt{2 \mathcal D(\xx,t)} \; \xi(\xx,t).  \label{eq:gGP}
\end{align}
Here $\gamma$ is the photon loss rate and $J$ the coupling between
the nearest-neighbor cavities \cite{kassenberg20,dung17}. {The
photons thermalize due to repeated absorption and emission by the
dye with respective rate coefficients $B_{12}$ and $B_{21}$.} The
ground (excited) molecular state occupation is denoted by $M_{1(2)}$
satisfying at all times $M_1(\xx)+M_2(\xx)=M$, where $M$ is the
number of dye molecules at each lattice site. The Kennard-Stepanov
relation \cite{kennard,stepanov,moroshkin14}  gives rise to energy
relaxation with dimensionless strength $\kappa= B_{12} \bar
M_1/(2T)$ (we set the Boltzmann constant $k_B=1$) \cite{gladilin20}.
The last term describes the spontaneous emission
noise~\cite{henry82,verstraelen19}: $ \mathcal{D}(\xx,t) = B_{21}
M_2(\xx,t) $ and {$\xi(\xx,t)$ is Gaussian white noise with
correlation function $\langle \xi(\xx,t) \xi(\xx',t') \rangle =
\delta_{\xx,\xx'} \delta(t-t')$.} The evolution of the number of
excited molecules due to interactions with the photons is opposite
to the change in number  of photons due to emission (both
deterministic and stochastic), absorption and energy relaxation. In
order to compensate for the loss of energy in the system, external
excitation with a pumping laser is needed.
Under the condition $J \ll T$, which assures that the occupations of
all momentum states are much larger than one, the generelized Gross
Pitaevskii classical field model \eqref{eq:gGP} is valid for all the
modes and there is no need to use a more refined quantum optical
approach \cite{kirton13,kirton15,kirton16}.

The noise in Eq. \eqref{eq:gGP} provides  a description of the
density and phase fluctuations. For the simplest case of a single
cavity, a crossover in the density fluctuations between a
`grandcanonical' regime with large fluctuations ($\delta n^2 \sim
\bar n^2$), for $\bar n^2\ll M_{\rm eff}$, and a `canonical' regime
with small fluctuations ($\delta n^2 \ll \bar n^2$), for $\bar
n^2\ll M_{\rm eff}$, have been observed \cite{klaers12,schmitt14}.
Here, the `effective' number of molecules is given by $\Meff =
(M+\gamma e^{- \Delta/T}/B_{21})/[2+2\cosh(\Delta/T)]$, where
$\Delta$ is the detuning between the cavity and the dye zero-phonon
transition frequency. For $e^{\Delta/T}\ll 1$, one has $M_{\rm eff}
\approx \bar M_2 \approx \eta M e^{\Delta/T}$ with $\eta=1+{\gamma
}/(2 \kappa T)$.

{\em Bogoliubov analysis --} While the linear Bogoliubov theory
\cite{bec_book,gladilin20} breaks down in the vicinity of the BKT
transition, that involves large phase differences between
neighbouring cavities, it nevertheless forms a good starting point
to obtain insight in the analytical dependence of the transition
temperature on the system parameters. From the linearized dynamics
of the density and phase fluctuations, one obtains the following
equation that relates the phase fluctuations to the density-phase
correlations \cite{gladilin20}
\begin{equation}
\kappa\langle |\delta \theta_\kk|^2 \rangle + \frac{1}{2 \bar n}
\langle \delta \theta_{-\kk} \delta n_{\kk} \rangle =
 \frac{\mathcal{D}_\theta}{\epsilon_\kk},
 \label{eq:bog_k}
\end{equation}
where the phase noise is
\begin{equation}
\mathcal{D}_\theta =
 \frac{B_{21}\bar M_2}{4\bar n}=\frac{\eta\kappa T}{2\bar n}
 \label{eq:d_theta}
\end{equation}
and the dispersion is of the tight binding form: $\epsilon_\kk = 2 J
[2- \cos(k_x)-\cos(k_y)]$.

{In equilibrium, invariance under time reversal ($\theta \rightarrow
-\theta$ and $\delta n \rightarrow \delta n$) ensures that the
second term in Eq. \eqref{eq:bog_k} vanishes.} In nonequilibrium
photon condensates, time reversal symmetry breaks down and the
density-phase correlator will play an important role in our
discussion of the phase fluctuations that lead to the BKT
transition.

At large momenta, the kinetic energy is much larger than the time
reversal breaking rates that involve the pumping and losses. The
density-phase correlations therefore become negligible at large $k$,
so that the phase fluctuations assume their thermal equilibrium
value. At low momenta on the other hand, the nonlinear dissipative
dynamics kicks in and deviations from the thermal behavior appear.
The crossover between nonlinear and ideal photon behavior occurs
around the momentum \cite{gladilin20,footnote2}
\begin{equation}
k_c  = \left[\frac{\gamma \kappa}{4J} \left(1+\frac{\bar n^2}{\bar
M_2}\right) \frac{\bar n^2}{\bar M_2}\right]^{1/6} \left(\frac{\eta
T}{J\bar n}\right)^{1/3}. \label{eq:kc}
\end{equation}
 At low temperatures, where phase fluctuations
are moderate, Eq. \eqref{eq:bog_k} is accurate for all momenta, but
close to the BKT temperature the linear approximation breaks down
and the system properties are determined by the full nonlinear
equations. Even then, at momenta $k > k_c$, the linear relation
\eqref{eq:bog_k} is expected to hold approximately.

In order to proceed further, we integrate \eqref{eq:bog_k} over all
momenta \cite{footnote1} to obtain for the local fluctuations:
\begin{equation}
\kappa \langle \delta \theta^2  \rangle + \frac{1}{2 \bar n} \langle
\delta \theta \, \delta n \rangle =
 \frac{\mathcal{D}_\theta}{2 \pi J} \left[c_1 + \ln(\pi/k_c)  \right],
 \label{eq:bog_x}
\end{equation}
where the constant $c_1$ approximates the contribution from the
momenta $k < k_c$, where \eqref{eq:bog_k} breaks down. It is clear
from the logarithmic dependence of \eqref{eq:bog_x} on $k_c$ that
phase ordering is impossible in the absence of dissipative
nonlinearity ($k_c \rightarrow 0$), reflecting the well known fact
that there is no phase transition for conservative noninteracting
bosons in two dimensions.

Since phase fluctuations at the BKT transition are large, the
parameter dependence of the transition point can be estimated
 with \eqref{eq:bog_x} by setting
 $\langle \delta \theta^2  \rangle \sim 1$, provided that
an estimate is available also for the the density-phase correlator.
In order to obtain a first approximation, we restrict temporarily to
one spatial dimension. Using partial integration to rewrite the
density-phase correlator as $ \langle \delta \theta \delta n \rangle
= L^{-1} \int dx \;
 \delta \theta \delta  n
= -  L^{-1}\int dx  \; (\partial \theta/\partial x) \delta N $,
where $\delta N = \int_0^x \delta n(x') dx'$, it can be related to
the current by use of the {identity $\partial j_x/\partial x  = -
\gamma \delta n$. This continuity relation shows that regions with
density suppression, such as a vortex core, behave as a source of
currents (see Fig. \ref{fig_curr})
\cite{wachtel,gladilin17,gladilin20b,aransonRMP02,staliunas}. With $j_x = 2 J \bar n (\partial
\theta/\partial x)$, this yields
\begin{equation}
\langle \delta \theta \, \delta n \rangle = \frac{1}{L}
\frac{\gamma}{2J \bar n}  \int dx \;\delta N^2(x) = \frac{\gamma}{2J
\bar n} \langle \delta N^2 \rangle.
\end{equation}

\begin{figure} \centering
\includegraphics[width=0.75 \linewidth]{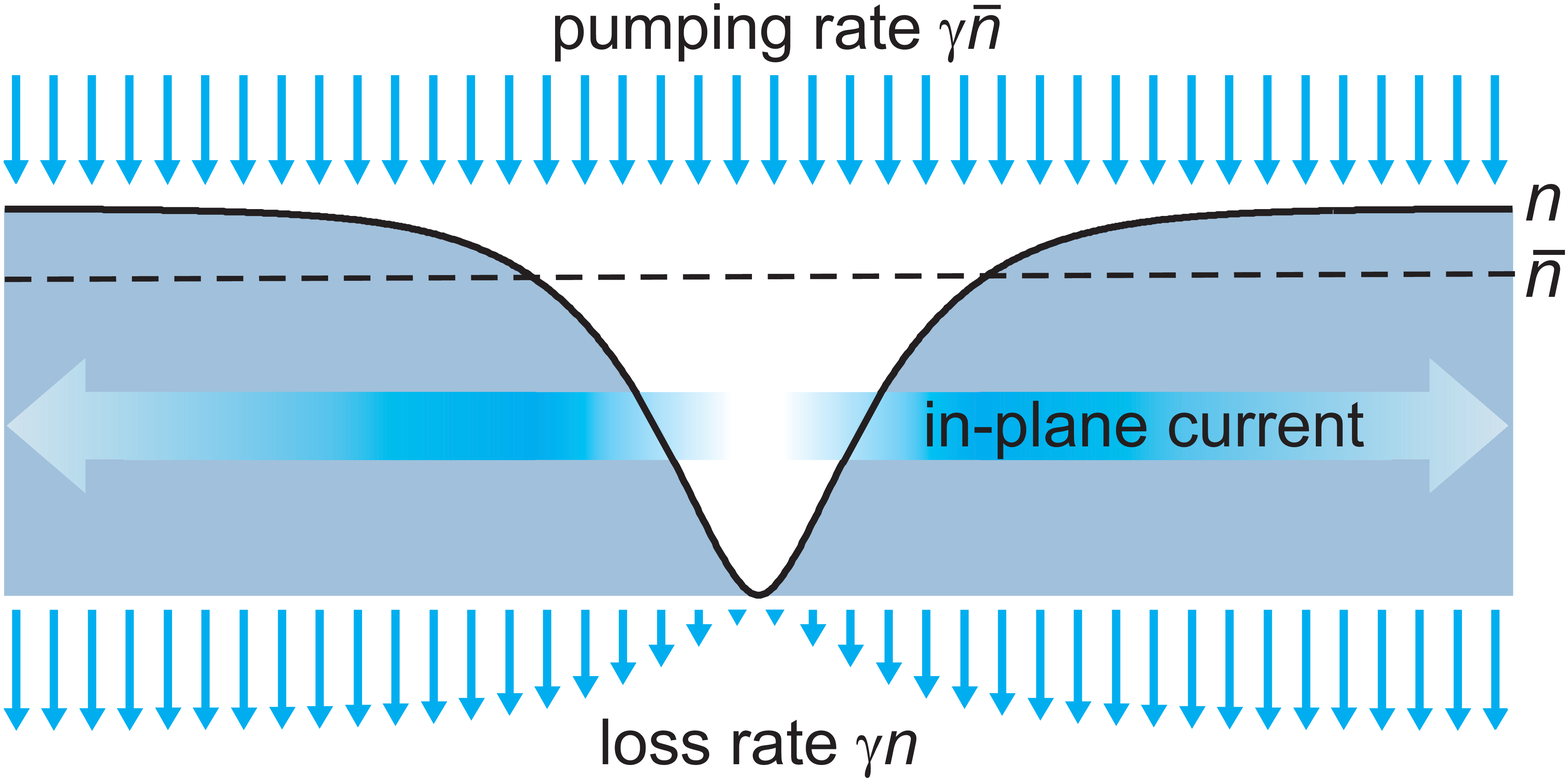}
\caption{Combination of a uniform pumping with {losses proportional
to the local density}
 leads to outgoing particle flows from
regions with reduced density. } \label{fig_curr}
\end{figure}

In order to estimate the expectation value $\langle \delta N^2
\rangle$ close to the transition, one can first consider a plane
density wave  of amplitude $a \bar n$, for which $ \langle
\delta N^2 \rangle \propto \bar n^2 a^2$.
At the transition, vortices have to nucleate, which requires in a
continuum model density fluctuations with amplitude $\bar n$ and
hence $a=1$. In a lattice geometry however, the vortex core can be
`in between' the lattice nodes and the density suppression smaller.
In Ref. \cite{gladilin20b}, the vortex core size was argued to
roughly behave as $r_0 \sim \sqrt{J/\gamma} $. Assuming an
exponential wave function in the vicinity of the vortex center, with
density profile $n(r) = \bar n \left[ 1- \exp(-r/r_0) \right]^2$,
one {can} estimate the minimal density modulation depth sufficient
to nucleate a vortex in the center of a plaquette as
\begin{equation}
a= 2 e^{-c_3 \sqrt{\gamma/J}} - e^{-2 c_3  \sqrt{\gamma/J}},
\label{eq:a}
\end{equation}
where a constant $c_3\sim 1$ was introduced. From the above
arguments, we come to the following estimate for the dependence of
the density-phase correlator on the system parameters close to the
transition point: $\langle \delta \theta \delta n \rangle = 2 c_2
\bar n a^2 \gamma/J$, with $c_2$ an additional fitting parameter.

{Using $\langle \delta \theta^2  \rangle = c_4 \sim1 $ together with
the above estimate of  $ \langle \delta \theta\; \delta n \rangle$}
allows to rewrite Eq. \eqref{eq:bog_x} as a relation for the
critical parameters:
\begin{equation}
\frac{J \bar n}{T} = \frac{\eta \kappa}{4\pi} \frac{c_1 + \ln
(\pi/k_c)} {c_4 \kappa + c_2 a^2 \gamma /J}, \label{eq:critrel}
\end{equation}
where $k_c$ and $a$ are given by Eqs. \eqref{eq:kc} and
\eqref{eq:a}. Because of the quite handwaiving arugments that have
led us to relation \eqref{eq:critrel}, it cannot be expected to hold
exactly. Nevertheless, we will show below that it offers a good
description of the critical point extracted from numerical
simulations for the following values of the fitting parameters:
 $c_1=3.56$, $c_2=0.132$,
$c_3=1.22$, $c_4=0.470$.

Figure \ref{fig_sum} shows the variations of the dimensionless
coupling constant $J \bar n/(\eta T)$ according to Eq.
\eqref{eq:critrel} as a function of the energy relaxation $\kappa$
and the ratio of losses to hopping $\gamma/J$ (note that both axes
are logarithmic) for three values of the number of photons per
cavity. We always restricted the dissipation strength to $\gamma/J
<2$ in order to keep a resolved photon dispersion.

Even though these parameters vary by orders of magnitude,
the variations in the coupling constant are quite moderate. For
large photon numbers (canonical regime, lower panel), the coupling
parameter is close to $0.6$ except for small $\kappa$, where
nonequilibrium effects are strongest. For smaller numbers of photons
(grandcanonical regime), a larger coupling constant is needed in
general and its variations are enhanced.

\begin{figure} \centering
\includegraphics[width=0.9 \linewidth]{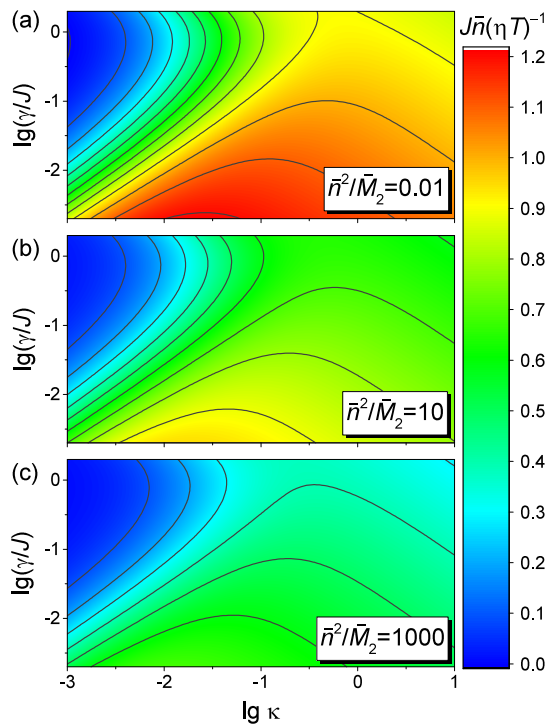}
\caption{Dimensionless coupling constant $J \bar n (\eta
T)^{-1}$, given by Eq. (\ref{eq:critrel}), as a function of $\kappa$
and $\gamma/J$ for three different values of $\bar n^2/\bar M_2$. }
\label{fig_sum}
\end{figure}

It is instructive to compare our relation \eqref{eq:critrel} to the
equilibrium BKT transition. The most elementary approximation is the
reduction of the lattice Bose gas to the XY model, obtained by
ignoring the density degree of freedom. The critical temperature has
been determined to be  $J \bar n /T = 0.56$ \cite{hsieh13}, which is
close to the value that we obtain in a large part of parameter space
in the deep canonical regime, see Fig. \ref{fig_sum}(c).

In the limit of small $\gamma$, the second term in the denominator
in Eq. (\ref{eq:critrel}), originating from the density-phase
correlator, can be neglected and we obtain for the critical coupling
parameter $J \bar n /T = \eta \ln(e^{c_1}\pi/k_c) /(4\pi c_4)$,
whose form is reminiscent of the transition point of the weakly
interacting Bose gas~\cite{prokofev01}. The main differences with
the equilibrium case are that the crossover momentum $k_c$ is now
determined by losses instead of interactions and the appearance of
the excess noise factor $\eta$. It is worth noting that in the limit
of small $\gamma$, where $\eta \rightarrow 1$, the prefactor of the
logarithm is in our fit equal to $1/(4 \pi c_4) \approx 1/(1.88 \,
\pi)$, close to the equilibrium prefactor of $1/(2 \pi)$
\cite{prokofev01}.

Formula \eqref{eq:critrel} is also reminiscent of the heuristic
expression $D_{\rm BKT}/n_{\rm BKT}\approx \kappa+0.003c$, which has
been shown to approximate fairly well the numerical results for the
critical noise-to-density ratio as a function of $\kappa$ and the
`nonequilibrium parameter' $c\propto \gamma$ in
interacting-polariton condensates~\cite{prb19}.

\textit{Numerical results --} Numerical simulations of the full gGPE
\eqref{eq:gGP} for an array of $100\times 100$ cavities with
periodic boundary conditions were done as explained in Ref.
\cite{gladilin20} and the location of the critical point was
determined as in Ref. \cite{prb19}: after a long time evolution in
the presence of noise, the system was evolved without noise for a
short time ($\sim 10$ ns) before checking for the presence of
vortices. This noiseless evolution gives the advantage of cleaning
up the photon phase while it is too short for the unbound
vortex-antivortex pairs to recombine. The propensity
for their recombination is reduced \cite{gladilin19} with respect to
the equilibrium case thanks to outgoing radial currents that provide
an effective repulsion between vortices and antivortices
\cite{wachtel}. If no vortices are present in the final photon
field, the system is said to be in the ordered phase; when vortex
pairs are present, it is denoted as disordered.

Because the presence of vortices and antivortices is susceptible to
statistical fluctuations, the numerical error on the transition
point is not only due to our finite steps in parameter values, but
also due to statistical uncertainty. The stochastic contribution to
the error bar is hard to quantify precisely, but the analysis of
many realizations of the dynamics allowed us to conclude that the
statistical error bars are typically not larger than the symbol
sizes in the figures.

As a first example of the parameter dependence of the critical
coupling, we show in Fig. \ref{fig_kappa} $J\bar n/T$ {as a function
of} $\kappa$, that was varied by changing both $M$ and $B_{21}$. The
numerically obtained results are indicated with the symbols (symbols
of the same type and color correspond to the same $M$ but different
$B_{21}$) and the fits with relation \eqref{eq:critrel} are shown
with full lines. Good correspondence is observed over the whole
range of $\kappa$, throughout which the critical coupling varies by
one order of magnitude. The initial rise and subsequent saturation
is clear from the explicit $\kappa$-dependence in Eq.
\eqref{eq:critrel}. As can be seen from the denominator in Eq.
\eqref{eq:critrel}, in the regime of small $\kappa$ the pumping and
losses, proportional to $\gamma$, are dominant. The reduction of the
critical coupling at small $\kappa$ can therefore be interpreted as
an increased robustness of the ordered phase due to driving and
dissipation, in analogy with Ref. \cite{prb19}. The decrease of
$J\bar n/T$ at large values of $\kappa$ originates from the
$k_c$-dependence on $\kappa$, while the increase with $M$ is due to
the dependence of $k_c$ on $\bar M_2\propto M$.

\begin{figure} \centering
\includegraphics[width=0.8 \linewidth]{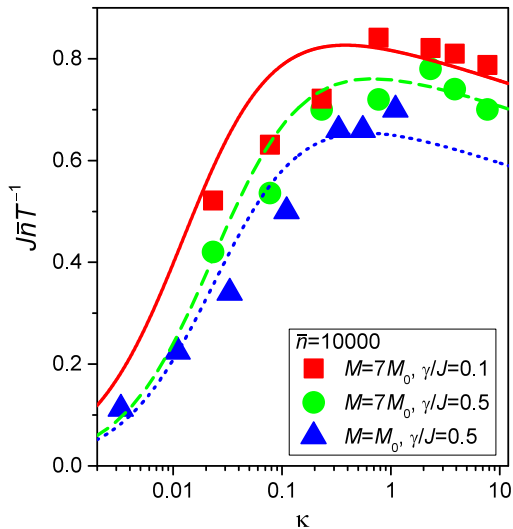}
\caption{Numerically determined dimensionless coupling strength
$J \bar n T^{-1}$  at the BKT transition (symbols) as a function of
$\kappa$ for different values of $M$ and $\gamma/J$. Here,
$M_0=10^9$, $\Delta/T = -5.2$. The curves correspond to Eq.
(\ref{eq:critrel}).} \label{fig_kappa}
\end{figure}

The dependence of the critical coupling on the number of photons per
lattice site is illustrated in Fig. \ref{fig_n}. Our
 relation \eqref{eq:critrel} again captures most of
the parameter dependence. According to Eq. \eqref{eq:critrel}, the
decrease of the critical coupling parameter with the number of
photons is due to the increase of the crossover momentum with
increasing $\bar n$. This trend is reflected in the numerical
results, but the numerical $\bar n$-dependence shows some features
that are not entirely reproduced by our formula and require a more
sophisticated theoretical approach.
At first sight, it could be surprising that a phase transition still
exists down to photon numbers as small as $\bar n=1000$, well in the
grandcanonical regime with $\bar n^2/M_{\rm eff} = 0.18$ and 0.026
for $M=10^9$ and $M=7\times 10^9$, respectively. Our explanation is
that for sufficiently strong coupling between the cavities, a subset
of $N$ cavities behaves collectively, thereby increasing the
effective $n^2/M_{\rm eff}$ linearly with $N$.

At $\gamma/J<1$ the critical coupling decreases with increasing
$\gamma/J$ (compare magenta to cyan or dark-cyan to orange
symbols/lines in Fig. \ref{fig_n}; see also Fig. \ref{fig_sum}). For
$\gamma/J\ll \kappa$ this decrease is determined by the behavior of
$\ln(\pi/k_c)$ in Eq. (\ref{eq:critrel}), while for larger
$\gamma/J$ the effect of the term $c_2 a^2\gamma/J$ dominates. In
the case of $\gamma/J\sim 1$, where the vortex core size is
comparable to the intercavity spacing and the density modulation
depth $a$ required for the BKT transition is significantly reduced,
the aforementioned decrease of the critical coupling with increasing
$\gamma/J$ can be fully canceled or even reversed due to a strong
decrease of $a^2$ in the term $c_2 a^2\gamma/J$ (compare green to
blue symbols/lines in Fig. \ref{fig_n}; see also Fig.
\ref{fig_sum}).

\begin{figure} \centering
\includegraphics[width=0.9 \linewidth]{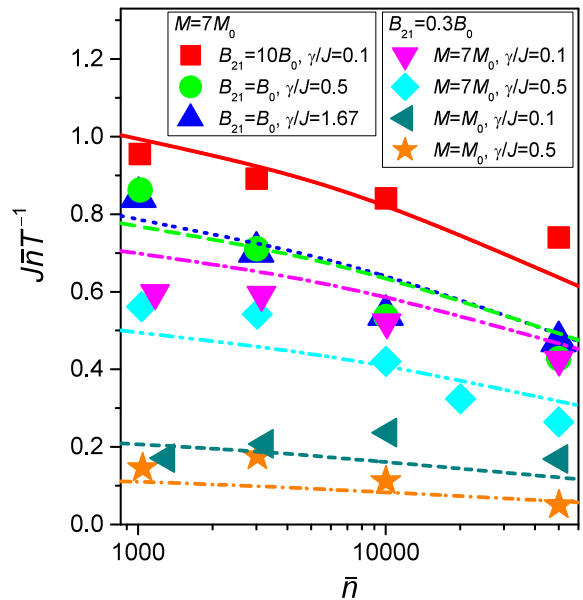}
\caption{Numerically determined dimensionless coupling strength
$J \bar n T^{-1}$  at the BKT transition (symbols) as a function of
$\bar n$ for different values of $M$, $B_{21}$ and $\gamma/J$.
Here, $M_0=10^9$, $B_0=10^{-7}$ meV, $\Delta/T = -5.2$. The
curves correspond to Eq. (\ref{eq:critrel}).} \label{fig_n}
\end{figure}

{\em Conclusions and outlook -- } Our numerical and analytical
investigations of a lattice nonequilibrium Bose-Einstein condensate
of noninteracting photons have shown that a BKT-like transition
exists between states with and without unbound vortex-antivortex pairs. According to
Eq. \eqref{eq:critrel}, the vortex-free phase is actually stabilized
by driving and dissipation. Our findings are in line with previous
numerical studies \cite{dagva15,caputo2016,prb19} of nonequilibrium
polariton condensates, where interactions were included.

The experimental verification of our prediction for the spontaneous
creation of vortices and antivortices should be possible by directly
measuring the phase profile of a photon condensate by
interferrometry as {used} for the observation of phase jumps of
localized photon condensates \cite{schmitt16}.

In our numerics, we have not found evidence for the destruction of
the ordered phase as was predicted on the basis of the description
of the phase dynamics by the nonlinear KPZ model
\cite{altman15,wachtel,sieberer}. This could be due to our finite
simulation area, but the interplay of BKT and KPZ physics in
nonequilibrium condensates \cite{squizzato18,gladilin14,he15} should
be explored further. The stability of the vortex free phase could be
important for potential applications to analog optical computations
\cite{berloff17,lagoudakis17,kalinin18,mcmahon16,yamamoto17,tradonsky19}
with photon condensates \cite{kassenberg20}.  From the side of
fundamental physics, it will also be interesting to study the
dynamics of a photon condensate lattice after a density quench
through the phase transition \cite{comaron18,kulczykowski17}.

{\em Acknowledgements --} We are grateful to Jan Klaers, Fahri
\"Ozt\"urk, Julian Schmitt, Martin Weitz, Iacopo Carusotto, Jacqueline Bloch and Wouter
Verstraelen for stimulating discussions.  VG was financially
supported by the the FWO-Vlaanderen through grant nr. G061820N.


\begin{references}
\bibitem{footnote2} The only difference with the expression in
Ref. \cite{gladilin20} is that there approximations  $1 \ll {\bar
n^2}/{\bar M_2}$ and $\eta \approx 1$ were made. These are relaxed
here.
\bibitem{footnote1} We approximate the tight binding lattice dispersion by its low energy quadratic behavior in order to obtain a simpler analytical expression. We have checked that using the full dispersion has only minor effects on the subsequent analysis.
\end{references}
\end{document}